\def\half{\frac{1}{2}}

\documentclass[amsmath,amssymb,showkeys, showpacs, superscriptaddress]{revtex4}
\usepackage{graphicx}
\usepackage{amsthm}
\usepackage{amsmath}
\usepackage{amssymb}
\usepackage{color}
\newtheorem{theorem}{Theorem}[section]

\theoremstyle{definition}

\parskip \baselineskip

\begin{document}

\hspace*{5 in}CUQM-149\
\vspace*{0.4 in}
\title{Spectra generated by a confined softcore Coulomb potential}
\author{Richard L. Hall}
\email{richard.hall@concordia.ca}
\affiliation{Department of Mathematics and Statistics, Concordia University,
1455 de Maisonneuve Boulevard West, Montr\'eal,
Qu\'ebec, Canada H3G 1M8}

\author{Nasser Saad}
\email{nsaad@upei.ca}
\affiliation{Department of Mathematics and Statistics,
University of Prince Edward Island, 550 University Avenue,
Charlottetown, PEI, Canada C1A 4P3.}

\begin{abstract}
\noindent  Analytic and approximate solutions for the energy eigenvalues generated by a confined softcore Coulomb potentials of the form $a/(r+\beta)$ in $d>1$ dimensions are constructed. The confinement is effected by linear and harmonic-oscillator potential terms, and also through `hard confinement' by means of an impenetrable spherical box. A byproduct of this work is the construction of polynomial solutions for a number of linear differential equations with polynomial coefficients, along  with the necessary and sufficient conditions for the existence of such solutions. Very accurate approximate solutions for the general problem with arbitrary potential parameters are found by use of the asymptotic iteration method.

\end{abstract}

\maketitle
\section{Introduction}\label{intro}
\subsection{Confined atoms in $d=3$ dimensions}
\noindent In order to fix ideas we considered first a simple model \cite{hall1981} for a soft confined atom obeying
a Schr\"odinger equation in $d=3$ dimensions of the form $(-\Delta +V)\psi = E\psi,$
 where $V(r)$ is an attractive central potential with Coulomb and confining terms.  If we assume a wave function of the form  
$\psi(r) = Y_{\ell}^{m}(\theta,\phi)r^{\ell}\exp(-g(r)),$ then we find the radial eigenequation  implies
\begin{equation}
(E-V(r))r = rg''(r)+2(l+1)g'(r) -r(g'(r))^2.
\end{equation}
If we now choose $g(r) = \half(vr+\omega r^2)$, $v>0$, $\omega >0,$ we obtain the 
 following family of exact solutions
\begin{equation}\label{modelsol}
V(r)=v\left(-\frac{\ell+1}{r} + \omega r\right) + \omega^2 r^2,\quad
 E = (3+2\ell)\omega - v^2/4,\quad l = 0,1,2,\dots
\end{equation}
The lowest radial excitations of the familiar Coulomb and oscillator problems are
 recovered from the special cases $\omega =0$ or $v = 0.$  Such specific exact solutions 
allow for analytical reasoning and explorations.  In addition, the explicit results provide relevant test
problems for the complementary approaches that must be used to complete the solution space.
 The question of the existence of exact solutions and the methods for finding them are
therefore an important part of the overall task.  As in the choice of the function $g(r)$ in
the simple illustration above, it is 
often the case in this context that exactness has something to do with polynomials.  Thus part of the paper involves the issue of
when an ordinary differential equation admits polynomial solutions.  As we shall see,  the `asymptotic iteration method' \cite{aim} plays a role both in the construction of exact solutions and also in finding approximations for arbitrary values of the problem parameters.
\subsection{Formulation of the problem in $d$ dimensions}
\noindent The Schr\"odinger equation in $d>1$ dimensions,
in atomic units $\hbar=2\mu=1$, with a spherically symmetric
potential $V(r)$ can be written as \cite{dong2011}
\begin{equation}\label{Sch_eq}
H\psi\equiv \left[-\Delta_d +V(r)\right]\psi({\bf r})=E\psi(\bf{r}),
\end{equation}
where $\Delta_d$ is the $d$-dimensional Laplacian operator and ${\bf r}=(x_1,x_2,\dots,x_d)$,  $r^2=\|{\bf r}\|^2=\sum_{i=1}^d x_i^2$. The quantum wave function $\psi$ is an element of the  Hilbert space $L^2(\Re^d).$  The principal class of spherically symmetric confining potentials we
 shall consider has the form
\begin{equation}\label{gn_pot}
V(r)=\dfrac{a}{r+\beta}+c\,r+b^2\,r^2, \quad \beta> 0, ~b>0.
\end{equation}
Thus $V(r)$ is continuous and $V(r)\rightarrow \infty$.  Consequently,  by Theorem XIII.67 of Reed-Simon-IV \cite{simon}, 
we know that $H$ has 
purely discrete eigenvalues and a complete set of eigenfunctions. Meanwhile, for $d \ge 3$, by Theorem XIII.69 of the same reference, we have a similar conclusion if we admit the Coulomb singularity in $V(r)$ by allowing $\beta = 0.$  For the case of hard confinement, $r\le R < \infty$ 
with Dirichlet boundary conditions at $r = R$,  
and $ \beta >0,$  so that $V(r)$ is continuous, we know from Theorem 23.56 of Ref.\cite{gs} that  again $H$ has a purely discrete spectrum.
These general results cover the cases we consider in this paper.
A comparable class of potentials  has been carefully analysed in Refs.\cite{bulla,alb}.  In order to express (\ref{Sch_eq}) in terms of 
$d$-dimensional spherical coordinates $(r, \theta_1, \theta_2,
\dots, \theta_{d-1})$, we separate variables using
\begin{equation}\label{gs_Sch_eq}
\psi({\bf r})=r^{-(d-1)/2}\,u(r)\, Y_{\ell_1,\dots,\ell_{d-1}}(\theta_1\dots\theta_{d-1}),
\end{equation}
where $Y_{\ell_1,\dots,\ell_{d-1}}(\theta_1\dots\theta_{d-1})$ is a
normalized spherical harmonic \cite{atkin} with characteristic value $\ell(\ell+d-2),$ and $\ell=\ell_1=0, 1, 2, \dots$ (the angular quantum numbers). One obtains the
radial Schr\"odinger equation as
\begin{equation}\label{gs_Sch_eq1}
\left[-{d^2\over dr^2}+{(k-1)(k-3)\over
4r^2}+V(r)-E\right] u_{n\ell}^{(d)}(r)=0, \quad\quad
\int_0^\infty \left\{u_{n\ell}^{(d)}(r)\right\}^2dr=1, ~~~
u_{n\ell}^{(d)}(0)=0,
\end{equation}
where $k=d+2\ell$.  Since the potential $V(r)$ is less singular than the centrifugal term,
$$u(r)\sim A\,r^{(k-1)/2},\qquad r\rightarrow 0,\qquad \mbox{where $A$ is a constant}.
$$
We note that the Hamiltonian and boundary
conditions of \eqref{gs_Sch_eq1} are invariant under the transformation
$$(d, \ell)\rightarrow (d\mp2, \ell\pm 1), $$
thus, given any solution for fixed $d$ and $\ell$, we can immediately
generate others for different values of $d$ and $\ell$. Further, the
energy is unchanged if $k=d+2\ell$ and the number of nodes $n$ is
constant. Repeated application of this transformation produces a
large collection of states; this has been discussed, for example, in Ref. \cite{doren1986}. 
\vskip0.1true in

\noindent In the present work we study the exact and approximate solutions of the Schr\"odinger eigenproblem generated by a confined soft-core Coulomb potential in $d$-dimensions, where $d>1.$  As we have discussed above, for the cases we consider, the spectrum of this problem is discrete, all
eigenvalues are real and simple, and they can be arranged in an
increasing sequence $\lambda_0<\lambda_1<\dots\rightarrow \infty$. The paper is organized as follows. In section \ref{sec2}, we set up the Schr\"odinger equation for the potential \eqref{gn_pot} and discuss the correspondence second-order differential equation.  In section \ref{method}, we present our method of solution that relies on the analysis of polynomial solutions of the differential equation
\begin{equation}\label{DE_general_sitting}
\left(\sum_{i=0}^{k}a_{k,i}\, r^{k-i} \right)y''+\left(\sum_{i=0}^{k-1}a_{k-1,i}\, r^{k-1-i} \right)y'-\left(\sum_{i=0}^{k-2}\tau_{k-2,i}\, r^{k-2-i} \right)y=0,\quad k\geq 2
\end{equation}
and different variants of this general differential-equation class. We discuss in particular necessary and sufficient conditions on the equation parameters for it to have polynomial solutions. A brief review of the asymptotic iteration method (AIM) is presented in section \ref{AIM}. In section \ref{app1}, the exact and approximate solutions for the problem are discussed, based on the results of section \ref{sec2}; and approximate solutions are found for arbitrary potential parameters $a,b,c$ and $\beta$ by an application of AIM . An analysis of the corresponding  exact and approximate solutions for the pure confined Coulomb case $\beta = 0$ is presented in section \ref{app2}. The `hard confinement' case, that is to say when the same system confined to the interior of an impenetrable spherical box of radius $R,$ is discussed in section \ref{spec}. In each of these sections, the results obtained are of two types: exact analytic results that are valid when certain
parametric constraints are satisfied, and accurate numerical values for arbitrary sets of potential parameters. 
\vskip0.1true in

\section{Setting up the differential equation}\label{sec2}

\noindent In this section, we consider the $d$-dimensional radial Schr\"odinger equation for $d>1$ :
\begin{equation}\label{Sch}
\left[-{d^2\over dr^2}+{(k-1)(k-3)\over 4r^2}+{a\over r+\beta}+c\,r+b^2r^2\right]u_{nl}^d(r)=E_{nl}^d u_{nl}^d(r),\quad \beta>0,\quad a\in(-\infty,\infty),\quad 0<r<\infty, \quad u(0) = 0.
\end{equation}
We note first that the differential equation (\ref{Sch}) has one regular singular
point at $r = 0$ with exponents given by the roots of the indicial equation 
\begin{equation}\label{indicial}
s(s-1)-{1\over 4}(k-1)(k-3)=0,
\end{equation}
and an irregular singular point at $r = \infty$.  For large $r$, the differential equation \eqref{Sch} assumes the asymptotic form
\begin{equation}\label{Sch_asy}
\left[-{d^2\over dr^2}+c\,r+b^2r^2\right]u_{nl}^d(r)\approx 0
\end{equation}
with an asymptotic solution 
\begin{equation}\label{asy_sol}
u_{nl}^d(r)\approx \exp\left(-\dfrac{c}{2b}\,r-\dfrac{b}{2}\,r^2\right).
\end{equation}
The roots $s$ of Eq.(\ref{indicial}), namely,
\begin{align*}
s_1={1\over 2}(3-k),\quad s_2={1\over 2}(k-1).
\end{align*}
determine the behaviour of $u_{nl}^d(r)$ as $r$ approaches $0$, only $s\geq 1/2$ is acceptable, since only in this case is the mean value of the kinetic energy finite \cite{landau}.
Thus, the exact solution of (\ref{Sch}) will assume the form
\begin{equation}\label{gen_sol}
u_{nl}^d(r)=r^{(k-1)/2}\exp\left(-\dfrac{c}{2b}\,r-\dfrac{b}{2}\,r^2\right)~f_n(r),\quad k=d+3l,
\end{equation}
where we note that $u_{nl}^d(r)\sim r^{(k-1)/2}$ as $r\rightarrow 0$. On  insertion of this ansatz wave function into (\ref{Sch}), we obtain the differential equation for $f_n(r)$ as
\begin{align}\label{secondorderde}
-4b^2\,r&\,(r+\beta)\,f_n''(r)+\left(8b^3\,r^3+4b\,(c+2b^2\,\beta)\,r^2+4b\,(b+c\beta-b\,k)\,r+4b^2\beta(1-k)\right)f_n'(r)
\notag\\
&+\left(\left[4b^2(b\,k-E)-c^2\right]\,r^2+(4 a b^2 - \beta c^2 - 4 b^2 \beta\,E + 2 b c (k-1) + 4 b^3 \beta k)\,r+2b\beta\,c(k-1)
\right)f_n(r)=0.
\end{align}
In the next section, we study the polynomial solutions of this differential equation which itself lies within a larger class of differential equations given by
\begin{align}\label{gen_DE}
(a_{4,2}r^2&+a_{4,3}r)y^{\prime \prime}+(a_{3,0}r^3+a_{3,1}r^2+a_{3,2}r+a_{3,3})y'-(\tau_{2,0}r^2+\tau_{2,1} r+\tau_{2,2})y=0,
\end{align}
where $\tau_{2,0},\tau_{2,1},\tau_{2,2}$ and $a_{i,j}$ are real constants for $i=3,4$ and $j=0,1,2,3$.
\vskip0.1true in
\section{The method of solution}\label{method}
\noindent The necessary condition  (\cite{h2010}, Theorem 6) for polynomial solutions $y(r)=\sum_{k=0}^n c_k r^k$ of the second-order linear differential equation  \eqref{gen_DE} is
\begin{equation}\label{poly_cond}
\tau_{2,0}=n\,a_{3,0},\quad n=0,1,2,\dots,
\end{equation}
provided $a_{3,0}^2+\tau_{2,0}^2\neq 0$.  The polynomial coefficients $c_n$ then satisfy the four-term recurrence relations 
\begin{align}\label{rec_rel}
((n-2)\, a_{3,0}-\tau_{2,0})\,c_{n-2}&+((n-1)\,a_{3,1}-\tau_{2,1})\,c_{n-1}+(n(n-1)\,a_{4,2}+ n\,a_{3,2}-\tau_{2,2})\,c_n\notag\\
&+(n(n+1)\,a_{4,3}+(n+1)\, a_{3,3} )\,c_{n+1}=0,\qquad c_{-2}=c_{-1}=0.
\end{align}
 The proof of \eqref{rec_rel} follows from an application of the  Frobenius method.
We note that the recurrence relations \eqref{rec_rel} can be written as a system of linear equations in the unknown coefficients $c_i$, $i=0,\dots, n$ given by
\begin{align}\label{lin_sys}
\begin{bmatrix}
\gamma_0&\delta_0&0&0&0&0 &\cdots&0&0 \\
\beta_1&\gamma_1&\delta_1&0&0&0&\cdots&0&0 \\
\alpha_2&\beta_2&\gamma_2&\delta_2&0&0&\cdots &0&0 \\
0&\alpha_3&\beta_3&\gamma_3&\delta_3&0&\cdots&0 &0 \\
\vdots & \vdots & \ddots & \ddots&\ddots&\ddots&\cdots&\vdots&\vdots\\
0&0&0&0&0&0 &\cdots&\beta_n&\gamma_n \\
\end{bmatrix}\begin{bmatrix}
c_0\\
c_1\\
c_2 \\
c_3 \\
\vdots\\
c_n \\
\end{bmatrix}=\begin{bmatrix}
0\\
0 \\
0 \\0 \\
\vdots\\
0 \\
\end{bmatrix}
\end{align}
where
\begin{align}\label{entries}
\gamma_n&=n(n-1)a_{4,2}+ na_{3,2}-\tau_{2,2},~\delta_n=n(n+1)a_{4,3}+(n+1) a_{3,3}, ~\beta_n= (n-1)a_{3,1}-\tau_{2,1},~
\alpha_n=(n-2) a_{3,0}-\tau_{2,0}.
\end{align}
Thus, for zero-degree polynomials $c_0\neq 0$ and $c_n=0,~n\geq 1$, we must have $\gamma_0=\beta_1=\alpha_2=0$, thus, in addition to the necessary condition $\tau_{2,0}=0$, the following two conditions become sufficient
\begin{equation}\label{poly0}
\tau_{2,2}=0,~\tau_{2,1}=0.
\end{equation} 
For the first-degree polynomial solution,  $ c_1\neq 0,$ and $c_n=0,~n\geq 2$, we must have 
$\gamma_0\,c_0+\delta_0\,c_1=0,\,\beta_1\,c_0+\gamma_1\,c_1=0,\,\alpha_2\,c_0+\beta_2\,c_2=0$ and $\alpha_3\,c_1=0$, thus, in addition to the necessary  condition $\alpha_3=0$  or
\begin{equation}\label{poly1_cond}
\tau_{2,0}=a_{3,0},
\end{equation}
it is also required that the following  two $2\times 2$-determinants simultaneously vanish
\begin{equation}\label{cond_poly1}
\left|\begin{array}{ccc}
-\tau_{2,2}&a_{3,3}\\
-\tau_{2,1}& a_{3,2} -\tau_{2,2}
\end{array}\right|=0,\quad
\mbox{and}\quad
\left|\begin{array}{ccc}
-\tau_{2,2}&a_{3,3}\\
-a_{3,0}& a_{3,1} -\tau_{2,1}
\end{array}\right|=0.
\end{equation}
For the second-degree polynomial solution, $c_2\neq0$ and $c_n=0$ for $n\geq 3$, it is necessary that $\gamma_0\,c_0+\delta_0\,c_1+\mu_0\,c_2=0,~\beta_1\,c_0+\gamma_1\,c_1+\delta_1\,c_2=0,
\alpha_2\,c_0+\beta_2\,c_1+\gamma_2\,c_2=0,\, \alpha_3\,c_1+\beta_3\,c_2=0,$ and $\alpha_4\,c_3=0$, from which we have the necessary condition
\begin{equation}\label{poly2_cond}
\tau_{2,0}=2\,a_{3,0}
\end{equation}
along with the  vanishing of the two $3\times 3$-determinants
\begin{equation}\label{cond_poly2}
\left|\begin{array}{ccc}
-\tau_{2,2}&a_{3,3}&0\\
-\tau_{2,1}& a_{3,2} -\tau_{2,2}&2a_{4,3}+ 2a_{3,3}\\
-2a_{3,0}&a_{3,1} -\tau_{2,1}&2a_{4,2}+2a_{3,2}-\tau_{2,2}
\end{array}\right|=0\quad\mbox{and}\quad
\left|\begin{array}{ccc}
-\tau_{2,2}&a_{3,3}&2a_{4,4}\\
-\tau_{2,1}& a_{3,2} -\tau_{2,2}&2a_{4,3}+ 2a_{3,3}\\
0&-a_{3,0}&2a_{3,1}-\tau_{2,1}
\end{array}\right|=0,
\end{equation} 
For the third-degree polynomial solution, $c_3\neq0$ and $c_n=0$ for $n\geq 4$,  we then have the necessary condition
\begin{equation}\label{cond_poly3}
\tau_{2,0}=3\,a_{3,0}
\end{equation}
along with the vanishing of the two $4\times 4$-determinants,
\begin{equation}\label{cond_poly3_1}
\left|\begin{array}{cccc}
-\tau_{2,2}&a_{3,3}&0&0\\
-\tau_{2,1}& a_{3,2} -\tau_{2,2}&2a_{4,3}+ 2a_{3,3}&0\\
-3a_{3,0}&a_{3,1} -\tau_{2,1}&2a_{4,2}+2a_{3,2}-\tau_{2,2}&3a_{3,3}+6a_{4,3}\\
0&-2a_{3,0} &2 a_{3,1}  - \tau_{2,1}&3 a_{3,2} + 6 a_{4,2} - \tau_{2,2}
\end{array}\right|=0
\end{equation} 
and
\begin{equation}\label{cond_poly3_2}
\left|\begin{array}{cccc}
-\tau_{2,2}&a_{3,3}&0&0\\
-\tau_{2,1}& a_{3,2} -\tau_{2,2}&2a_{4,3}+ 2a_{3,3}&0\\
-3a_{3,0}&a_{3,1} -\tau_{2,1}&2a_{4,2}+2a_{3,2}-\tau_{2,2}&3a_{3,3}+6a_{4,3}\\
0&0&-a_{3,0} &3a_{3,1} - \tau_{2,1}
\end{array}\right|=0
\end{equation} 
For the fourth-degree polynomial solution ($n=4$), $c_4\neq 0$ and $c_n=0$ for $n\geq 5$,  we then have the necessary condition
\begin{equation}\label{cond_poly3}
\tau_{2,0}=4\,a_{3,0}
\end{equation}
along with the vanishing of the two $5\times 5$-determinants,
\begin{equation}\label{cond_poly3_1}
\left|\begin{array}{ccccc}
-\tau_{2,2}&a_{3,3}&0&0&0\\
-\tau_{2,1}& a_{3,2} -\tau_{2,2}&2a_{4,3}+ 2a_{3,3}&0&0\\
-4a_{3,0}&a_{3,1} -\tau_{2,1}&2a_{4,2}+2a_{3,2}-\tau_{2,2}&3a_{3,3}+6a_{4,3}&0\\
0&-3a_{3,0} &2 a_{3,1} - \tau_{2,1}&3 a_{3,2} + 6 a_{4,2} - \tau_{2,2}&4a_{3,3}+12a_{4,3}\\
0&0&-2 a_{3,0}&3 a_{3,1}- \tau_{2,1}& 4 a_{3,2} + 12 a_{4,2} - \tau_{2,2}
\end{array}\right|=0
\end{equation} 
and
\begin{equation}\label{cond_poly3_2}
\left|\begin{array}{ccccc}
-\tau_{2,2}&a_{3,3}&2a_{4,4}&0&0\\
-\tau_{2,1}& a_{3,2} -\tau_{2,2}&2a_{4,3}+ 2a_{3,3}&0&0\\
-4a_{3,0}&a_{3,1} -\tau_{2,1}&2a_{4,2}+2a_{3,2}-\tau_{2,2}&3a_{3,3}+6a_{4,3}&0\\
0&-3a_{3,0} &2 a_{3,1} - \tau_{2,1}&3 a_{3,2} + 6 a_{4,2} - \tau_{2,2}&4a_{3,3}+12a_{4,3}\\
0&0&0&- a_{3,0} &4 a_{3,1} - \tau_{2,1}
\end{array}\right|=0
\end{equation} 
 Similar expressions for higher-order polynomial solutions can be easily obtained. The vanishing of these determinants can be regarded as the sufficient conditions under which the coefficients $\tau_{2,1}$ and $\tau_{2,2}$ of Eq. \eqref{secondorderde} can be expressed  in terms of the other parameters.

\vskip0.1true in
\section{The asymptotic iteration method and some related results}\label{AIM}
\noindent The asymptotic iteration method (AIM) is an iterative algorithm originally
introduced \cite{aim} to investigate the analytic and approximate solutions of the differential
equation
\begin{equation}\label{AIM_Eq}
y''=\lambda_0(r) y'+s_0(r) y,\quad\quad ({}^\prime={d\over dr})
\end{equation}
where $\lambda_0(r)$ and $s_0(r)$ are $C^{\infty}-$differentiable
functions. A key feature of this method is to note the invariant
structure of the right-hand side of (\ref{AIM_Eq}) under further
differentiation. Indeed, if we differentiate \eqref{AIM_Eq} with
respect to $r$, we obtain
\begin{equation}\label{first_der}
y^{\prime\prime\prime}=\lambda_1(r)\, y^\prime+s_1(r)\, y
\end{equation}
where $\lambda_1= \lambda_0^\prime+s_0+\lambda_0^2$ and
$s_1=s_0^\prime+s_0\lambda_0.$ Further differentiation of
equation \eqref{first_der}, we obtain
\begin{equation}\label{second_der}
y^{(4)}=\lambda_2(r)\, y^\prime+s_2(r)\, y
\end{equation}
where $\lambda_2= \lambda_1^\prime+s_1+\lambda_0\lambda_1$ and
$s_2=s_1^\prime+s_0\lambda_1.$ Thus, for $(n+1)^{th}$ and
$(n+2)^{th}$ derivative of (\ref{AIM_Eq}), $n=1,2,\dots$, we have
\begin{equation}\label{the_nth_der}
y^{(n+1)}=\lambda_{n-1}(r)\,y^\prime+s_{n-1}(r)\, y
\end{equation}
and
\begin{equation}\label{the_np1th_der}
y^{(n+2)}=\lambda_{n}(r)\,y^\prime+s_{n}(r)\,y
\end{equation}
respectively, where
\begin{equation}\label{AIM_seq}
\lambda_{n}=
\lambda_{n-1}^\prime+s_{n-1}+\lambda_0\lambda_{n-1}\hbox{ ~~and~~
} s_{n}=s_{n-1}^\prime+s_0\lambda_{n-1}.
\end{equation}
From \eqref{the_nth_der} and \eqref{the_np1th_der}, we have
\begin{equation}\label{poly_Co}
\lambda_n y^{(n+1)}- \lambda_{n-1}y^{(n+2)} = \delta_ny {\rm
~~~where~~~}\delta_n=\lambda_n s_{n-1}-\lambda_{n-1}s_n.
\end{equation}
Clearly, from \eqref{poly_Co} if $y$, the solution of
(\ref{AIM_Eq}), is a polynomial of degree $n$, then $\delta_n\equiv
0$. Further, if $\delta_n=0$, then $\delta_{n'}=0$ for all $n'\geq
n$. In an earlier paper \cite{aim}, we proved the principal theorem
of AIM, namely 
\vskip0.1in

\begin{theorem}  Given $\lambda_0$ and $s_0$ in
$C^{\infty}(a,b),$ the differential equation (\ref{AIM_Eq}) has the
general solution
\begin{equation}\label{AIM_solution}
y(r)= \exp\left(-\int\limits^{r}{s_{n-1}(t)\over \lambda_{n-1}(t)} dt\right) \left[C_2
+C_1\int\limits^{r}\exp\left(\int\limits^{t}(\lambda_0(\tau) +
2{s_{n-1}\over \lambda_{n-1}}(\tau)) d\tau \right)dt\right]
\end{equation}
if for some $n>0$
\begin{equation}\label{tm_cond}
\delta_n=\lambda_n s_{n-1}-\lambda_{n-1}s_n=0.
\end{equation}
where $\lambda_n$ and $s_n$ are given by \eqref{AIM_seq}.
\end{theorem}

\vskip0.1true in
\noindent Recently, it has been shown \cite{aim1} that the termination condition \eqref{tm_cond} is necessary and sufficient for the differential
equation \eqref{AIM_Eq} to have polynomial-type solutions of degree at most $n$, as we may conclude from Eq.\eqref{poly_Co}. The application of AIM to a number of problems has been outlined in many publications. The applicability  of the method is not restricted to a particular class of differentiable functions (e.g. polynomials or rational functions), rather, it can accommodate any type of differentiable function. The fast convergence of the iterative scheme depend on a suitable  choice for the starting values of $r=r_0$ and the correct asymptotic solutions near the boundaries \cite{saad2008}.
\section{Exact and approximate solutions for the soft-confined softcore Coulomb potential}\label{app1}
\begin{figure}[!h]
\centering
\includegraphics[scale=1.2]{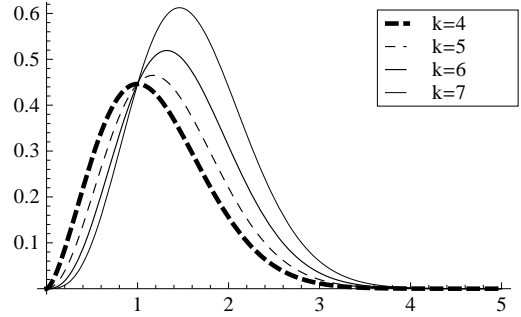}
\caption{Un-normalized ground state wave functions as given by \eqref{wf1_gen_sol} for specific values of $b=\beta=1$ and different values of $k$.}\label{Fig1}
\end{figure}

\noindent Comparing equation \eqref{secondorderde} with \eqref{gen_DE} and using parameters given by
\begin{align}\label{parameters}
a_{4,2}&=-4b^2,\qquad a_{4,3}=-4b^2\beta, \notag\\
a_{3,0}&=8b^3,\qquad a_{3,1}=4b\,(c+2b^2\,\beta),\qquad a_{3,2}=4b\,(b+c\beta-b\,k),\qquad a_{3,3}=4b^2\beta(1-k),\notag\\
\tau_{2,0}&=c^2+4b^2(E_{nl}^d-b\,k),\qquad \tau_{2,1}=-4 a b^2 + \beta c^2 + 4 b^2 \beta\,E_{nl}^d - 2 b c (k-1) - 4 b^3 \beta k,\qquad \tau_{2,2}=-2b\beta\,c(k-1),
\end{align}
the exact solution of \eqref{Sch} assumes the following form
\begin{equation}\label{gen_sol_exp}
u_{n\ell}^d(r)=r^{(k-1)/2}\exp\left(-\dfrac{c}{2b}\,r-\dfrac{b}{2}\,r^2\right)~\sum_{i=0}^{n'} {C}_i\,r^i,\quad k=d+3l,
\end{equation}
where $f_{n'}(r)=\sum_{i=0}^{n'} {C}_i\,r^i$ and  $n$ counts the number of zeros of $f_{n'}(r)=0$, hence the number of nodes in the wave function solution. The coefficients $ {C}_i$ can be easily evaluated using the four-term recurrence relations \eqref{rec_rel},
\begin{align}\label{rec_relation}
4 b^2 (i-2 - n') {C}_{i-2} &+ (2 a b + c ( k -3+ 2 i) + 4 b^2 \beta ( i-1 - n'))  {C}_{i-1} + 
(\beta c (k-1 + 2 i)-2 b i (k-2 + i)) {C}_i \notag\\
&- 
    2 b \beta (i+1) (k-1 + i)  {C}_{i+1}=0,\quad  {C}_{-1}=0,~~ {C}_0=1,~~ {C}_1=c/(2b),~~i\geq 2,
\end{align}
using the necessary condition 
\begin{equation}\label{gen_poly_cond}
E_{n\ell}^d=b\,(2n'+k)-\dfrac{c^2}{4b^2},\quad n'=0,1,2,\dots.
\end{equation}
The potential parameters $a,b,c$ and $\beta$ satisfy 
sufficient conditions according to the following scenarios:  for a zero-degree polynomial solution, $n'=0$, if $f_0(r)=1$, the ground-state solution of equation \eqref{Sch} is given by
\begin{equation}\label{zero_poly_sol}
u_{0\ell}^d(r)=r^{(k-1)/2}\exp\left(-\dfrac{c}{2b}\,r-\dfrac{b}{2}\,r^2\right),
\end{equation}
with ground-state eigenenergy  
\begin{equation}\label{ground_state_energy}
E_{0\ell}^d=b\,k-\dfrac{c^2}{4b^2}\quad\mbox{subject to the parameter conditions}\quad c=\dfrac{2ab}{1-k}\quad\mbox{and}\quad \beta=0.
\end{equation}
In the next section, we shall focus on the case of $\beta=0$ which corresponds to the Coulomb potential perturbed by an added polynomial in $r$. In the rest of this section, we shall assume $\beta>0.$  For a first-degree polynomial solution, $n'=1$, 
 \begin{equation}\label{first_poly_sol}
f_0(r)=1+\dfrac{c}{2b}\,r,
\end{equation}
and the exact solution wave function of equation \eqref{Sch} reads
\begin{equation}\label{wf1_gen_sol}
u_{0\ell}^d(r)=r^{(k-1)/2}\left(1+\dfrac{c}{2b}\,r\right)\exp\left(-\dfrac{c}{2b}\,r-\dfrac{b}{2}\,r^2\right),\qquad  E_{0\ell}^d=b\,(k+2)-\dfrac{c^2}{4b^2},
\end{equation}
subject to the following two conditions related the potential parameters
\begin{equation}\label{cond1and2}
 4a\,b^2+\beta\,(c^2(k+1)-8b^3)=0\quad\mbox{and}\quad 2abc+c^2(k+1)-8b^3=0.\
\end{equation}
Since, by assumption $b>0$ and $\beta>0$, $c=2b/\beta >0$ and the polynomial solution $f_i(r)=f_0(r)$ has no roots, in which case  $E_{0\ell}^d$ represent a ground-state solution of the Schr\"odinger equation \eqref{Sch} subject to the parameters $a$, $b$ and $c$ satisfying the conditions given by \eqref{cond1and2}. In summary, the exact solutions of Schr\"odinger's equation
\begin{equation}\label{Sch_Exact_1}
\left[-{d^2\over dr^2}+\dfrac{(k-1)(k-3)}{4r^2}+{2b\beta^2-(k+1)\over \beta(r+\beta)}+\dfrac{2b}{\beta}\,r+b^2r^2\right]u_{0\ell}^d(r)=E_{0\ell}^d u_{0\ell}^d(r),\quad \beta>0,\quad 0<r<\infty.
\end{equation}
is explicitly given by
\begin{equation}\label{case1_gen_sol}
u_{0\ell}^d(r)=r^{(k-1)/2}\left(1+\dfrac{r}{\beta}\right)\exp\left(-\dfrac{r}{\beta}-\dfrac{b}{2}\,r^2\right),\qquad  E_{0l}^d=b\,(k+2)-\dfrac{1}{\beta^2},
\end{equation}
In Figure \ref{Fig1}, we display the un-normalized 
ground-state solution using  \eqref{case1_gen_sol} for $b=\beta=1$ and different values of $k$. 
For the rest of the spectrum we use the asymptotic iteration method as described in section \ref{AIM}, starting with
\begin{equation}\label{aim_gen_sol}
u_{nl}^d(r)=r^{(k-1)/2}\left(1+\dfrac{c}{2b}\,r\right)\exp\left(-\dfrac{c}{2b}\,r-\dfrac{b}{2}\,r^2\right)\,g(r)
\end{equation}
where $g(r)=1$ corresponds to the exact solution \eqref{wf1_gen_sol}, equation \eqref{Sch_Exact_1} yields the second-order differential for $g(r)$ as
\begin{equation}\label{dE_g}
g''(r)=\left(\dfrac{2}{\beta} + \dfrac{1 - k}{r} + 2 b r -\dfrac{2}{\beta + r}\right)g'(r)+\left(2b-E+b\,k-\dfrac{1}{\beta^2}\right)g(r).
\end{equation}
Hence, we may initiate AIM with
\begin{equation}\label{aim_start}
\lambda_0(r)=\dfrac{2}{\beta} + \dfrac{1 - k}{r} + 2 b r -\dfrac{2}{\beta + r}\qquad\mbox{and}\qquad
  s_0(r)= 2b-E+b\,k-\dfrac{1}{\beta^2}
\end{equation} 
The question is then to find the initial value $r_0$ that stabilizes the computation of the termination-condition roots \eqref{tm_cond}. To this end, we take the highest of the absolute values among all the roots of 
$$V(r)-E_{0\ell}^d=  \dfrac{(k-1)(k-3)}{4r^2}+{2b\beta^2-(k+1)\over \beta(r+\beta)}+\dfrac{2b}{\beta}\,r+b^2r^2 - \left(b\,(k+2)-\dfrac{1}{\beta^2}\right)=0$$ 
which yields $r_0\sim 3$, henceforth  we shall fix $r_0$ at $r_0 = 3$ for all of our numerical computations. In Table \ref{table:tab1}, we report our results from AIM for first 12 decimal places. The eigenvalue reported in Table \ref{table:tab1}  were computed using Maple version 16 running on an IBM architecture personal computer and we have chosen a high-precision environment. In order to accelerate our computation we have written our own code for a root-finding algorithm instead of using the default procedure {\tt Solve} of \emph{Maple 16}. The results of AIM may be obtained to any degree of precision, although we have reported our results to only the first twelve decimal places, 

\begin{table}[!h] \caption{The first few Eigenenergies $E_{n0}^{d=4}$ of Schr\"odinger equation \eqref{Sch_Exact_1}. The initial value utilize AIM is $r_0=3$. The subscript $N$ refers to the number of iteration used by AIM.\\ } 
\centering 
\begin{tabular}{||c |p{2.2in}||c| p{2.2in}|}
\hline
$n$&$E_{n0}^{d=4}$&$n$&$E_{n0}^{d=5}$\\ \hline
$0$&$~5.000~000~000~000_{N=3,Exact}$&$0$&$~6.000~000~000~000_{N=3,Exact}$\\ \hline
$1$&$10.223~655~148~231_{N=90}$&$1$&$11.139~009~555~512_{N=81}$\\ \hline
$2$&$15.140~755~138~866_{N=91}$&$2$&$16.025~939~658~710_{N=83}$\\ \hline
$3$&$19.899~975~543~589_{N=92}$&$3$&$20.771~696~017~356_{N=85}$\\ \hline
$4$&$24.559~997~330~221_{N=92}$&$4$&$25.425~173~414~020_{N=110}$\\ \hline
$5$&$29.150~691~578~737_{N=118}$&$4$&$30.012~690~909~013_{N=109}$\\ 
 \hline
\end{tabular}
\label{table:tab1}
\end{table}
\vskip0.1true in

\begin{figure}[ht]
\centering
\includegraphics[scale=1.2]{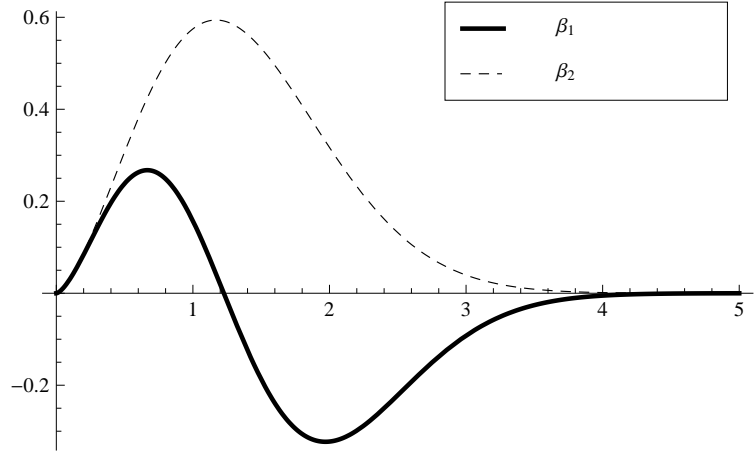}
\caption{Un-normalized ground-state and first-excited wave functions as given by \eqref{wf2_gen_sol_2} for $k=4, b=c=1$ and specific values of $\beta$.}\label{wfcase2}
\end{figure}

\noindent For a second-degree polynomial solution, $n'=2$, of \eqref{secondorderde}, we have
 \begin{equation}\label{second_poly_sol}
f_i(r)=1+\dfrac{c}{2b}\,r+\dfrac{4ab^2+\beta((k+1)c^2-16b^3)}{8\,b^2\beta\, k}\, r^2,
\end{equation}
and the exact solution of equation \eqref{Sch} reads
\begin{equation}\label{wf2_gen_sol}
u_{i\ell}^d(r)=r^{(k-1)/2}\left(1+\dfrac{c}{2b}\,r+\dfrac{4ab^2+\beta((k+1)c^2-16b^3)}{8\,b^2\beta\, k}\, r^2\right)\exp\left(-\dfrac{c}{2b}\,r-\dfrac{b}{2}\,r^2\right),\qquad  E_{i\ell}^d=b\,(k+4)-\dfrac{c^2}{4b^2},
\end{equation}
where $i$ counts the number of roots of the polynomial solution \eqref{second_poly_sol} subject to the simultaneous conditions relating the parameters $a,~b,~c$ and $\beta$, 
 \begin{align}\label{second_poly_cond}
4 a b^2 (-4 b k + 3 \beta c (1 + k)) +  \beta^2 c (c^2 (1 + k) (3 + k) - 16 b^3 (3 + 2 k))&=0,\notag\\
8 a^2 b^3 + 2 a b (-16 b^3 \beta + \beta c^2 (1 + k) + 2 b c (3 + k)) + 
 \beta c (c^2 (1 + k) (3 + k) - 16 b^3 (3 + 2 k))&=0.
\end{align}
In particular, for
$$a=4 b \beta-\frac{\beta c^2}{4 b^2}+\left(\frac{c }{b}-\frac{2 }{\beta}-\frac{\beta c^2}{4 b^2}\right)\,k$$
The exact solution of the Schr\"odinger equation for $0<r<\infty$
\begin{equation}\label{Sch_Exact_2}
\left[-{d^2\over dr^2}+\dfrac{(k-1)(k-3)}{4r^2}+\dfrac{4 b \beta-\frac{\beta c^2}{4 b^2}+\left(\frac{c }{b}-\frac{2 }{\beta}-\frac{\beta c^2}{4 b^2}\right)\,k}{r+\beta}+c\,r+b^2r^2\right]u_{il}^d(r)=\left(b\,(k+4)-\dfrac{c^2}{4b^2}\right)u_{il}^d(r)
\end{equation}
is
\begin{equation}\label{wf2_gen_sol_2}
u_{il}^d(r)=r^{(k-1)/2}\left(1+\dfrac{c}{2b}\,r+\frac{\beta\,c-2 b}{2 b \beta^2}\, r^2\right)\exp\left(-\dfrac{c}{2b}\,r-\dfrac{b}{2}\,r^2\right)
\end{equation}
subject to the relation among the parameters $b$, $c$ and $\beta$ given by
\begin{equation}\label{root_eq}
-16 b^3 k+4 b^2 \beta c (3+5 k)+b\,\beta^2 \left(32 b^3-8 c^2 (1+k)\right)+c\,\beta^3 \left(c^2 (1+k)-8 b^3\right)=0
\end{equation}
As an example, for $b=c=1$ and $k$, the roots of equation \eqref{root_eq} for $\beta>0$ are
$\beta_1=0.760~237~519~523~249~5$ and $\beta_2=3.854~071~917~077~363~6$. We display in Figure \ref{wfcase2} the exact solutions as given by \eqref{wf2_gen_sol_2}.
\noindent For a third-degree polynomial solution, $n'=3$, of equation \eqref{secondorderde}, 
 \begin{equation}\label{third_poly_sol}
f_i(r)=1+\dfrac{c}{2b}\,r+\dfrac{4ab^2+\beta((k+1)c^2-24b^3)}{8\,b^2\beta\, k}\,  r^2+\dfrac{4 a b^2 (3 \beta c (1+k)-4 b k)+\beta^2 c \left(c^2 (1+k) (3+k)-8 b^3 (9+7 k)\right)}{48 b^3\beta^2 k (1+k)}\,r^3,
\end{equation}
and the exact solution of equation \eqref{Sch} reads
\begin{align}\label{wf2_gen_sol}
u_{i\ell}^d(r)&=r^{(k-1)/2}\exp\left(-\dfrac{c}{2b}\,r-\dfrac{b}{2}\,r^2\right)\notag\\
&\left(1+\dfrac{c}{2b}\,r+\dfrac{4ab^2+\beta((k+1)c^2-24b^3)}{8\,b^2\beta\, k}\,  r^2+\dfrac{4 a b^2 (3 \beta c (1+k)-4 b k)+\beta^2 c \left(c^2 (1+k) (3+k)-8 b^3 (9+7 k)\right)}{48 b^3\beta^2 k (1+k)}\,r^3\right),
\end{align}
where
\begin{equation}\label{third_poly_En}
E_{il}^d=b\,(k+6)-\dfrac{c^2}{4b^2},
\end{equation}
and the parameters $a,b,c$ and $\beta$ satisfy, by means of \eqref{cond_poly3_1} and \eqref{cond_poly3_2}, the conditions
 \begin{align}\label{third_poly_cond}
48\, a^2\, b^4\, \beta\, (1 + k) &- 
 8\, a\, b^2\, (48\, b^3\, \beta^2\, (1 + k) - 12\, b^2\, k\, (1 + k) + 
    8\, b\, \beta\, c\, k\, (2 + k) - 3\, \beta^2\, c^2\, (1 + k)\, (3 + k)) \notag\\
    &+ 
 \beta^3\, (576\, b^6\, (1 + k) + c^4\, (1 + k)\, (3 + k)\, (5 + k) - 
    16\, b^3\, c^2\, (24 + 5\, k\, (5 + k)))=0\notag\\ \notag\\
    8\, a^2\, b^3\, (-4\, b\, k &+ 3\, \beta\, c\, (1 + k)) + 
 2 \,a\, b\, (\left[\beta^2\, c^3\, (3 + k)-48\, b^4 \beta\right]\, (1 + k) - 
    8\, b^2\, c\, k\, (5 + k) + 6\, b\, \beta\, c^2\, (1 + k)\, (5 + k) \notag\\
    &- 
    8\, b^3\, \beta^2\, c\, (9 + 7 k)) + 
 \beta^2\, (\left[576\, b^6 + c^4\, (3 + k)\, (5 + k)\right](1+k) - 
    16\, b^3\, c^2\, (24 + 5\, k\, (5 + k)))=0.
\end{align}
\vskip0.1true in
\noindent For arbitrary values of the potential parameters $a$, $b$, $c$ and $\beta$ that do not necessarily obey the above conditions, we may use AIM directly to compute the eigenvalues \emph{accurately},  as the zeros of the termination condition (\ref{tm_cond}). The method above can also be used to verify the exact solutions we have obtained earlier. For arbitrary parameters, we employ  AIM with
\begin{equation}\label{aim_start}
\lambda_0(r)=\dfrac{1-k}{r}+2\,b\,r+\dfrac{c}{b}\qquad\mbox{and}\qquad 
  s_0(r)= \dfrac{c(k-1)}{2\,b\,r}+\dfrac{a}{r+\beta}+\dfrac{4b^3\,k-c^2-4b^2E_{nl}^d}{4b^2}.
\end{equation} 
and compute the AIM sequences $\lambda_n$ and $s_n$  as given by Eq.(\ref{AIM_seq}). We note that for given values of the potential parameters $a$, $b,$ and of $k=d+2\ell$, the termination condition  $\delta_n=\lambda_n s_{n-1}-\lambda_{n-1}s_n=0
$ yields again an expression that depends on both $r$ and $E$. Thus, in order to use AIM as an approximation technique for computing the eigenvalues $E$ we need to feed AIM with a suitable initial value of $r=r_0$ that could stabilize AIM (that is, to avoid oscillations). Again, for our calculations in Table \ref{table:tab2}, we have used $r_0=3$. 

\begin{table}[!h] \caption{Eigenvalues $E_{0\ell}^{d=4,5}$ for $V(r)=1/(r+1)+r+r^2$ and different values of the angular momentum $\ell$. The initial value used by AIM is $r_0=3$. The subscript $N$ refers to the number of iteration used by AIM.\\ } 
\centering 
\begin{tabular}{||c |p{3.6in}||c| p{3.6in}|}
\hline
$\ell$&$E_{0\ell}^{d=4}$&$l$&$E_{0\ell}^{d=5}$\\ \hline
$0$&$~5.743~064~598~822_{N=75}=E_{01}^{2}$&$0$&$~6.881~699~763~857_{N=69}=E_{01}^{3}$\\ \hline
$1$&$~8.010~441~473~733_{N=63}=E_{02}^{2}=E_{00}^{6}$&$1$&$~9.131~165~616~720_{N=56}=E_{02}^{3}=E_{00}^{7}$\\ \hline
$2$&$10.245~221~261~283_{N=52}=E_{03}^{2}=E_{01}^{6}=E_{00}^{8}$&$2$&$11.353~616~525~901_{N=48}=E_{03}^{3}=E_{01}^{7}=E_{00}^{9}$\\ \hline
$3$&$12.457~128~050~974_{N=44}=E_{04}^{2}=E_{02}^{6}=E_{01}^{8}=E_{00}^{10}$&$3$&$13.556~369~149~873_{N=41}=E_{04}^{3}=E_{02}^{7}=E_{01}^{9}=E_{00}^{11}$\\ \hline
$4$&$14.651~834~191~239_{N=39}=E_{05}^{2}=E_{03}^{6}=E_{02}^{8}=E_{01}^{10}=E_{00}^{12}$&$4$&$15.743~928~601~250_{N=38}=E_{05}^{3}=E_{03}^{7}=E_{02}^{9}=E_{01}^{11}=E_{00}^{13}$\\ \hline
$5$&$16.832~989~791~994_{N=35}=E_{06}^{2}=E_{04}^{6}=E_{03}^{8}=E_{02}^{10}=E_{01}^{12}=E_{00}^{14}$&$5$&$17.919~302~156~447_{N=32}=E_{06}^{3}=E_{04}^{7}=E_{03}^{9}=E_{02}^{11}=E_{01}^{13}=E_{00}^{15}$\\ \hline
\hline
\end{tabular}
\label{table:tab2}
\end{table}

\section{Exact and approximate solutions for the pure Coulomb
potential plus linear and oscillator radial terms}\label{app2}

\noindent In this section, we focus our attention on the case of $\beta=0$, specifically we study the exact and approximate eigenenergies of a hydrogenic atom with a Coulomb potential \cite{castro} in the presence of an external linear term and an harmonic oscillator. We have 
\begin{align}\label{Pert_coul}
V(r)=\dfrac{a}{r}+c\,r+b^2\,r^2,\qquad a\neq 0,\quad b>0.
\end{align}
This soft-confined potential has been the subject of intensive study over the past few decades in a wide range of contexts \cite{roy1988,bessis1987,roy1990,hall2004,hall1996}. In the light of solutions to the equation \eqref{gen_DE}, we discuss the quasi-exact solutions of Schr\"odinger equation for the potential \eqref{Pert_coul} and their connection with the solution of the biconfluent heun equation \cite{ron1995,dutra,leaute1986, leaute1990,arriola1991,ovsiyuk, caruso} where we extend the some of the known results to arbitrary dimensions and provide a compact analytic solutions that we use to verify our approximation method using AIM. For this purpose, we set $\beta=0$ in the differential equation \eqref{secondorderde} to obtain
\begin{align}\label{Pert_coul_1}
r&\,f_n''(r)+\left(-2b\,r^2-\dfrac{c}{b}\,r+k-1\right)f_n'(r)
+\left(\left(E-b\,k+\dfrac{c^2}{4b^2}\right)\, r-a+\dfrac{(1-k)\,c}{2b}
\right)f_n(r)=0.
\end{align}
This equation can easily be compared with  \eqref{gen_DE} for $a_{4,3}=a_{3,3}=\tau_{3,3}=0$, in which case equation \eqref{gen_DE} reduces to
\begin{align}\label{gen_DE_sp}
a_{4,2}r& y^{\prime \prime}+(a_{3,0}r^2+a_{3,1}r+a_{3,2})y'+(-\tau_{2,0}r-\tau_{2,1})y=0,
\end{align}
with polynomial solutions $y_n=\sum_{j=0}^n  {C}_j\, r^j$ only if $\tau_{2,0}=n\,a_{3,0},$ and polynomial coefficients $ {C}_j$ that satisfy the three-term recurrence relations, derived by use of the Frobenius method,
given by 
\begin{align}\label{rec_rel_sp}
((n-1) a_{3,0}-\tau_{2,0}) {C}_{n-1}+(na_{3,1}-\tau_{2,1}) {C}_{n}&+(n+1)(na_{4,2}+ a_{3,2}) {C}_{n+1}=0,\qquad  {C}_{-1}=0,\quad  {C}_0=1.
\end{align}
In this case, the first few polynomials are 
\begin{align*}
f_0(r)&=1\qquad{providing}\quad \tau_{2,1}=0,\\
f_1(r)&=
1+\frac{\tau_{2,1}}{a_{3,2}}\, r\qquad{providing}\quad
\left|\begin{array}{ccc}
-\tau_{2,1}&a_{3,2}\\
-a_{3,0}& a_{3,1} -\tau_{2,1}
\end{array}\right|=0,\\
f_2(r)&=1+\dfrac{\tau_{2,1}}{a_{3,2}}\, r+\dfrac{
2a_{3,0}a_{3,2}-a_{3,1}\tau_{2,1}+\tau_{2,1}^2}{2a_{3,2}(a_{3,2}+a_{4,2})}\,r^2 \qquad{providing}\quad\left|\begin{array}{cccc}
-\tau_{2,1} &a_{3,2} &0\\
-2 a_{3,0}& a_{3,1} - \tau_{2,1}& 2 (a_{3,2} + a_{4,2})\\
0&-a_{3,0}&2 a_{3,1} - \tau_{2,1}
\end{array}\right|=0,\\
f_3(r)&=1+\dfrac{\tau_{2,1}}{a_{3,2}}\, r+\dfrac{
3a_{3,0}a_{3,2}-a_{3,1}\tau_{2,1}+\tau_{2,1}^2}{2a_{3,2}(a_{3,2}+a_{4,2})}\,r^2 +\dfrac{a_{3,0} (-6 a_{3,1} a_{3,2} + 7 a_{3,2} \tau_{2,1} + 4 a_{4,2} \tau_{2,1}) + 
 \tau_{2,1} (2 a_{3,1}^2 - 3 a_{3,1} \tau_{2,1} + \tau_{2,1}^2)}{6 a_{3,2} (a_{3,2} + a_{4,2}) (a_{3,2} + 2 a_{4,2})}\,r^3,\notag\\
&\qquad{providing}\quad\left|\begin{array}{ccccc}
-\tau_{2,1} &a_{3,2} &0&0\\
-3 a_{3,0} & a_{3,1} - \tau_{2,1} & 2 a_{3,2} + 2 a_{4,2}&0\\
0&-2 a_{3,0}& 2 a_{3,1} - \tau_{2,1}&3 a_{3,2} + 6 a_{4,2}\\
0&0&-a_{3,0}& 3 a_{3,1} - \tau_{2,1}\\
\end{array}\right|=0.
\end{align*}
On other hand, as noted earlier,  \eqref{Pert_coul_1} is a special case of the biconfluent Heun differential equation \cite{ron1995,rovder,hautot1969}
\begin{equation}\label{bheun}
zf^{\prime\prime}(z) +(1+\alpha-\beta z-2z^2)f^\prime(z)+\left[(\gamma-\alpha-2)\,z-\dfrac12(\delta+(1+\alpha)\beta)\right]f(z)=0.
\end{equation}
Indeed, by a simple comparison, with $z=\sqrt{b}\,r$, between \eqref{bheun} and \eqref{Pert_coul}, we find by using
\begin{equation}\label{eq50}
\alpha=k-2,\quad \beta=\dfrac{c}{b^{3/2}},\quad \gamma=\dfrac{E}{b}+\dfrac{c}{4b^3},\qquad \delta = \dfrac{2a}{\sqrt{b}}.
\end{equation}
that we can express the analytic solutions of \eqref{gen_DE_sp} in terms of the  Bi-confluent Heun functions \cite{ron1995,rovder,hautot1969} as
\begin{align}\label{bheun_func}
f(r)=H_B\left(k-2,\dfrac{c}{b^{3/2}},\dfrac{E}{b}+\dfrac{c}{4b^3}, \dfrac{2a}{\sqrt{b}},\sqrt{b}\, r\right)
\end{align}
with polynomial solutions providing $E_{nl}^d=b\,(2n'+k)-{c^2}/{4b^2},n'=0,1,2,\dots$. To this end, the polynomial solutions of the differential equation
\begin{align}\label{Pert_coul_2}
r&\,f_{n'}''(r)+\left(-2b\,r^2-\dfrac{c}{b}\,r+k-1\right)f_{n'}'(r)
+\left(2\,b\,n'\, r-a+\dfrac{(1-k)\,c}{2b}
\right)f_{n'}(r)=0,\qquad n'=0,1,2,\dots
\end{align}
are
\begin{align}\label{bheun_func}
f_{n'}(r)=H_B\left(k-2,{c}\,{b^{-3/2}},2\,n'+k, 2\,a\,b^{-1/2},\sqrt{b}\, r\right)=\sum_{j=0}^{n'}  {C}_j\, r^j
,\qquad n'=0,1,2,\dots.
\end{align}
where the coefficients $ {C}_j$ are easily computed by means of the three-term recurrence relations, using \eqref{rec_rel_sp},
\begin{align}\label{Re-Rel}
(j + 1)(j + k - 1)\, {C}_{j+1} + \left(\dfrac{(1 - k)c}{2b} - a -\dfrac{c}{b}j\right)  {C}_j + 
  2b(n' - j + 1)\,  {C}_{j-1}= 0,\quad   {C}_{-1} = 0,\quad  {C}_0 = 1,
  \end{align}
subject to the termination condition $  {C}_{j+1}=0.$
Thus, the first few polynomial solutions are given explicitly as
\begin{align}
&f_0(r)=1,\qquad{providing }\qquad 2ab+(k-1)\,c=0,\label{case1}\\ \notag\\
&f_1(r)=1+\dfrac{2ab+(k-1)\,c}{2b(k-1)}\,r,\qquad{providing}\qquad 8b^3(1-k)+4b^2a^2+4kcba+c^2(k^2-1)=0,\label{case2}\\ \notag\\
&f_2(r)=1+\dfrac{2ab+(k-1)\,c}{2b(k-1)}\,r+\dfrac{16b^3(1-k)+4b^2a^2+4kcba+c^2(k^2-1)}{8b^2\,k\,(k-1)}\, r^2,\qquad{providing}\notag\\
&32a(1-2k)b^4+8 \left(a^3 - 2 c (-1 + k) (3 + 2 k)\right)\,b^3+12 a^2 c (1 + k)\,b^2-2 a c^2 (1 - 3 k (2 + k))\,b+c^3(k^2-1)(k+3)=0. \label{case3}
\end{align}
For arbitrary values of the potential parameters,  we may initiate the asymptotic iteration method to solve the eigenvalue problem independently of the above mentioned constraints. Although, AIM was applied previously to study this potential  \cite{barakat,amore}, we claim that we obtain here more accurate and consistent numerical results.  Using AIM with
\begin{equation}\label{aim_start}
\lambda_0(r)=\dfrac{1-k}{r}+2\,b\,r+\dfrac{c}{b} \qquad\mbox{and}\qquad  s_0(r)= \dfrac{c(k-1)+2ab}{2\,b\,r}+\dfrac{4b^3\,k-c^2-4b^2E_{nl}^d}{4b^2},
\end{equation} 
and computing the AIM sequences $\lambda_n$ and $s_n$ using \eqref{AIM_seq}, we evaluate, recursively, the roots of the termination condition \eqref{tm_cond}, starting with the initial value $r_0=3$, similar to the technique used to report Table \ref{table:tab2}. In Table \ref{table:difdn3}, we use AIM to verify the `exact' ground state energy \eqref{case1} for $a=b=1$, then apply AIM to the higher excited states. In Table \ref{table:difdn4}, it is clear that we have greatly improved on the earlier AIM results of Barakat \cite{barakat}. These results also highlight the conclusion obtained by Amore et. al. \cite{amore} on the fast convergence of AIM  for this particular problem. In Table \ref{table:difdn5} using the Riccati-Pad\'e method (RPM), we report a simple comparison comparing our results with those  obtained earlier by Amore et. al. \cite{amore}. An immediate reason for the improvement noted in the results of Tables \ref{table:difdn4} and \ref{table:difdn5} a consequence of the appropriate structures of the asymptotic solutions near zero and infinity \eqref{gen_sol_exp}. This illustrates the importance of using a more adequate asymptotic solution \cite{saad2008} that usually yields better stability, convergence, and accuracy of AIM. 

\begin{table}[h] \caption{Eigenvalues $E_{n0}^{d=3,4,5,6}$ for $V(r)=1/r+cr+r^2$ where $c$ is determined from \eqref{case1}. The initial value used by AIM is $r_0=3$. The subscript $N$ refers to the number of iteration used by AIM.\\ } 
\centering 
\begin{tabular}{|c|c |p{2.5in}||c|c| p{2.5in}|}
\hline
$c$&$n$&$E_{n0}^{d=3}$&$c$&$n$&$E_{n0}^{d=4}$\\ \hline
$-1$&0&$~2.750~000~000~000~000~000_{N=3,exact}$&$-2/3$&0&$~3.888~888~888~888~888~889_{N=3~Exact}$\\
~&1&$~6.105~909~691~182~920~708_{N=64}$&~&1&$~7.485~841~099~550~171~275_{N=59}$\\
~&2&$~9.615~295~284~487~204~826_{N=62}$&~&2&$11.169~992~576~098~137~834_{N=58}$\\
~&3&$13.210~469~278~706~047~371_{N=61}$&~&3& $14.905~199~749~925~709~834_{N=57}$ \\
~&4&$16.860~555~849~138~091~010_{N=59}$&~&4&$18.674~207~558~484~831~292_{N=57}$\\
~&5&$20.549~102~541~238~464~811_{N=57}$&~&5&$22.467~438~445~946~572~167_{N=55}$\\
~&6&$24.266~299~867~653~311~177_{N=59}$&~&6&$26.279~004~288~368~339~228_{N=54}$\\ \hline
\hline
$c$&$n$&$E_{n0}^{d=5}$&$c$&$n$&$E_{n0}^{d=6}$\\ \hline
$-1/2$&0&$~4.937~500~000~000~000~000_{N=3,exact}$&$-2/5$&0&$~5.960~000~000~000~000~000_{N=3~Exact}$\\
~&1&$~8.655~823~170~124~162~086_{N=56}$&~&1&$~9.749~149~491~375~024~656_{N=50}$\\
~&2&$12.428~555~074~489~786~355_{N=54}$&~&2&$13.574~797~401~850~504~632_{N=50}$\\
~&3&$16.234~977~694~977~922~106_{N=52}$&~&3& $17.424~191~007~631~759~307_{N=49}$ \\
~&4&$20.064~504~343~130~534~075_{N=52}$&~&4&$21.290~390~825~325~282~344_{N=48}$\\
~&5&$23.910~981~048~253~203~499_{N=53}$&~&5&$25.169~188~410~054~967~854_{N=48}$\\
~&6&$27.770~499~635~352~076~648_{N=50}$&~&6&$29.057~829~460~632~247~615_{N=48}$\\ \hline
\hline
\end{tabular}
\label{table:difdn3}
\end{table}

\begin{table}[!h] \caption{A comparison between selected eigenenergies calculated by Barakat \cite{barakat} and in the present work.\\ } 
\centering 
\begin{tabular}{|c|c|c|c|c|p{2.0in}| p{0.5in}|}
\hline
$n$&$\ell$&$a$&$c$&$b$&$E_{nl}^{3}$&$\epsilon_{Barakat}$\\ \hline
$0$&$0$&$-2$&$0.894~42$&$\sqrt{0.2}$&$0.341~633~800~749~479~644_{N=45}$& $0.341~64$\\ \hline
$0$&$1$&$-2$&$0.447~22$&$\sqrt{0.2}$&$1.986~079~419~684~181~694_{N=33}$&$1.986~06 $\\ \hline
$0$&$2$&$-2$&$0.298~14$&$\sqrt{0.2}$&$3.019~378~385~388~245~576_{N=30}$& $3.019~38 $\\ \hline
$0$&$3$&$-2$&$0.223~60$&$\sqrt{0.2}$&$3.962~403~145~424~275~957_{N=37}$& $3.962~42 $\\ \hline
 \hline
$0$&$0$&$-2$&$8.944~28$&$2\sqrt{5}$&$12.416~411~447~380~603~566_{N=83}$& $~6.208~20$\\ \hline
$0$&$1$&$-2$&$4.472~14$&$2\sqrt{5}$&$22.110~682~447~511~408~897_{N=72}$&$22.110~64 $\\ \hline
$0$&$2$&$-2$&$2.981~42$&$2\sqrt{5}$&$31.193~837~323~750~444~679_{N=64}$& $31.193~86 $\\ \hline
$0$&$3$&$-2$&$2.236~06$&$2\sqrt{5}$&$40.186~716~024~771~681~149_{N=60}$& $20.093~36 $\\ \hline
 \hline
\end{tabular}
\label{table:difdn4}
\end{table}

\begin{table}[!h] \caption{A comparison between eigenenergies as obtained by Amore et. al. \cite{amore} using Riccati–Pad\'e method (RPD) and those of the present work $E_{AIM}$, for particular values of the parameter $c$ in the potential $V(r)=-2/r+c\,r+\sqrt{2}\, r^2$.\\ } 
\centering 
\begin{tabular}{|c|p{2.0in}|p{2.4in}|}
\hline
$c$&$E_{RPM}$&$E_{AIM}$\\ \hline
$-4$&$-2.343~347~169~439~4$& $-2.343~347~169~439~302~087~596~937_{N=93}$\\ \hline
$-2$&$-0.452~373750~381~743~8$& $-0.452~373~750~381~743~858~907~206_{N=89}$\\ \hline
$~~2$&$~~2.665~690~984~529~681~669~8$& $~~2.665~690~984~529~681~669~856~944_{N=74}$\\ \hline
$~~4$&$~~4.029~812~452~923~474~112~0$& $~~4.029~812~452~923~474~111~929~868_{N=66}$\\ \hline
 \hline
\end{tabular}
\label{table:difdn5}
\end{table}

\section{Exact and approximate solutions with hard confinement $r \le R$.}\label{spec}
\noindent In this section, we turn our  attention to study the $d$-dimensional radial Schr\"odinger equation 
\begin{equation}\label{Sch_conf}
\left[-{d^2\over dr^2}+{(k-1)(k-3)\over 4r^2}+V(r)\right]u_{nl}^d(r)=E_{nl}^d\, u_{nl}^d(r),\qquad \int_0^R |u_{nl}^d(r)|^2dr<\infty, ~~~
u_{n\ell}^{(d)}(0)=u_{n\ell}^{(d)}(R)=0,
\end{equation}
with the potential
\begin{equation}\label{pot_conf}
V(r)=\left\{ \begin{array}{ll}
\dfrac{a}{r+\beta}+c\,r+b^2r^2, &\mbox{ if\quad $0<r< R$}, \\ \\
  \infty, &\mbox{ if\quad $r\geq  R$,}
       \end{array} \right.
\end{equation}
where $u_{nl}^d(0)=u_{nl}^d(R)=0$. We employ the following ansatz for the wave function
\begin{equation}\label{wf_conf}
u_{nl}^d(r)=r^{(k-1)/2}\,(R-r)\,\exp\left(-\dfrac{c}{2b}\,r-\dfrac{b}{2}\,r^2\right)f_n(r),\quad k=d+2l,
\end{equation}
where  the $(R-r)$ factor is inserted to ensure the vanishing of the radial wave function $u_{nl}^d(r)$ at the boundary $r=R$. On substituting \eqref{wf_conf} into \eqref{Sch_conf}, we obtain the following second-order differential equation for the functions $f_n(r)$,
\begin{align}\label{DE_conf}
\bigg(&-4 b^2 r^3 + 4 b^2 (R-\beta)\, r^2 + 4 b^2 \beta R\, r\bigg)f_n''(r)+\bigg(8 b^3 r^4 +4 b (c + 2 b^2 (\beta - R)) \,r^3-4 b (b - \beta c + b k + 2 b^2 \beta R + c R)\,r^2\notag\\
&-4 b (\beta c R + b (\beta + \beta k + R - k R))\, r + 4\, b^2 \beta\, (k-1)\, R\bigg)f_n'(r)\notag\\
&+\bigg[\left( 4 b^3 (2 + k)-c^2 - 4 b^2 E \right)\,r^3+(4 a b^2 + 2 b c (1 + k) + (c^2 + 4 b^2 E) (R-\beta) + 
 4 b^3 (\beta (k+2) - k R))\,r^2\notag\\
&+(2 b\, (\beta\, c\, (1 + k)-2 b\, (k-1) )+(\beta\, (c^2 + 4 b^2 E)-4 a b^2  - 2 b c (k-1) - 4 b^3 \beta k)R)\,r-2 b \beta (k-1) (2 b + c R)\bigg]f_n(r)=0.
\end{align}
This differential equation goes beyond the equation discussed in section \ref{method}, so we introduce another more general class of differential equation that that allows us to analyze the polynomial solutions of \eqref{DE_conf}.
\begin{theorem}\label{thmVII.1}
The second-order linear differential equation 
\begin{align}\label{DE543}
(a_{5,0}\, r^5+ a_{5,1}\, r^4 +  a_{5,2}\, r^3 + a_{5,3}\, r^2 + a_{5,4}\, r +  a_{5,5} )
  f''(r)&+(a_{4,0} r^4+ + a_{4,1} r^3  + a_{4,2} r^2  + a_{4,3} r +a_{4,4})f'(r)\notag\\
  &-(\tau_{3,0} r^3 +\tau_{3,1}\,r^2 +\tau_{3,2}\,r+\tau_{3,3})f(r)=0
\end{align}
has a polynomial solution 
$y(r)=\sum_{j=0}^nc_j r^j$,
if
\begin{equation}\label{DEE543_Nec_cond}
\tau_{3,0}=n\,(n-1)\,a_{5,0}+n\,a_{4,0},\qquad n=0,1,2,\dots,
\end{equation}
provided $a_{5,0}^2+a_{4,0}^2+\tau_{3,0}^2\neq 0$. The polynomial coefficients $c_n$ then satisfy the following six-term recurrence relation
\begin{align}\label{Rec_conf}
((j-3)&(j-4)a_{5,0}+(j-3) a_{4,0}-\tau_{3,0})\,c_{j-3}+((j-2)(j-3)a_{5,1}+ (j-2)a_{4,1}-\tau_{3,1})\,c_{j-2}\notag\\
&+((j-1)(j-2)a_{5,2}+ (j-1)a_{4,2}-\tau_{3,2})\,c_{j-1}+(j(j-1)a_{5,3}+j a_{4,3}-\tau_{3,3} )\,c_{j}\notag\\
&+(j(j+1)a_{5,4}+(j+1)a_{4,4})\,c_{j+1}+(j+1)(j+2)\,a_{5,5}\,c_{j+2}=0
\end{align}
with
$c_{-3}=c_{-2}=c_{-1}=0$. In particular, for the zero-degree polynomials $f_0(r)=1$ where $c_0=1$ and $c_n=0,~n\geq 1$, we must have $\tau_{3,0}=0$ along with  
\begin{equation}\label{Case1_conf}
\tau_{3,1}=0,\quad\tau_{3,2}=0,\quad\tau_{3,3}=0.
\end{equation} 
For the first-degree polynomial solution $$ f_1(r)=1+\dfrac{\tau_{3,3}}{a_{4,4}}\, r,$$  where $c_0=1,~ c_1=\tau_{3,3}/a_{4,4},$ and $c_n=0,n\geq 2$, we must have $
\tau_{3,0}=a_{4,0}
$
along with the vanishing of the three $2\times 2$-determinants, simultaneously,
\begin{equation}\label{Case2_conf}
\left|\begin{array}{ccc}
-\tau_{3,3}&a_{4,4}\\
-\tau_{3,2}& a_{4,3} -\tau_{3,3}
\end{array}\right|=0,\qquad
\left|\begin{array}{ccc}
-\tau_{3,3}&a_{4,4}\\
-\tau_{3,1}& a_{4,2} -\tau_{3,2}
\end{array}\right|=0,\qquad
\mbox{and}\qquad
\left|\begin{array}{ccc}
-\tau_{3,3}&a_{4,4}\\
-a_{4,0}& a_{4,1} -\tau_{3,1}
\end{array}\right|=0.
\end{equation}
For the second-degree polynomial solution, $$f_2(r)=1+\frac{(a_{4,4}+a_{5,4})\tau_{3,3}-a_{5,5} \tau_{3,2}}{a_{4,4} (a_{4,4}+a_{5,4})+a_{5,5} (\tau_{3,3}-a_{4,3})}\,r+\frac{a_{4,4}\tau_{3,2}+\tau_{3,3} (\tau_{3,3}-a_{4,3})}{2 (a_{4,4} (a_{4,4}+a_{5,4})+a_{5,5} (\tau_{3,3}-a_{4,3}))}\,r^2$$ where $c_n=0$ for $n\geq 3$, we must have
$
\tau_{3,0}=2a_{5,0}+2a_{4,0}$ 
along with the vanishing of the three $3\times 3$-determinants, simultaneously,
\begin{align}\label{Case3_conf}
&\left|\begin{array}{ccc}
-\tau_{3,3}&a_{4,4}&2a_{5,5}\\
-\tau_{3,2}& a_{4,3} -\tau_{3,3}&2a_{5,4}+ 2a_{4,4}\\
-\tau_{3,1}&a_{4,2} -\tau_{3,2}&2a_{5,3}+2a_{4,3}-\tau_{3,3}
\end{array}\right|=0,\quad
\left|\begin{array}{ccc}
-\tau_{3,3}&a_{4,4}&2a_{5,5}\\
-\tau_{3,2}& a_{4,3} -\tau_{3,3}&2a_{5,4}+ 2a_{4,4}\\
-2a_{5,0}-2a_{4,0}&a_{4,1} -\tau_{3,1}&2a_{5,2}+2a_{4,2}-\tau_{3,2}
\end{array}\right|=0,
\end{align}
and
\begin{align}
\left|\begin{array}{ccc}
-\tau_{3,3}&a_{4,4}&2a_{5,5}\\
-\tau_{3,2}& a_{4,3} -\tau_{3,3}&2a_{5,4}+ 2a_{4,4}\\
0&-2a_{5,0}-a_{4,0}&2a_{5,1} +2a_{4,1}-\tau_{3,1}
\end{array}\right|=0,
\end{align} 
For third-degree polynomial solution,
\begin{align*}
&f_3(r)=1+\dfrac{2a_{5,5}^2 \tau_{3,1}+(a_{4,4}+a_{5,4}) (a_{4,4}+2a_{5,4})\tau_{3,3}+a_{5,5} \left(\tau_{3,3}^2-2 (a_{4,3}+a_{5,3}) \tau_{3,3}-(a_{4,4}+2a_{5,4}) \tau_{3,2}\right)}{a_{4,4}^3+3a_{4,4}^2a_{5,4}+2a_{5,5} (-a_{4,3}a_{5,4}+a_{4,2} a_{5,5}-a_{5,5}\tau_{3,2}+a_{5,4}\tau_{3,3})+a_{4,4} \left(2a_{5,4}^2+a_{5,5} (2 \tau_{3,3}-3a_{4,3}-2a_{5,3})\right)}\,r\notag\\
&+\dfrac{a_{4,4}^2\tau_{3,2}+2\tau_{3,3} (-a_{4,3}a_{5,4}+a_{4,2}a_{5,5}-a_{5,5}\tau_{3,2}+a_{5,4}\tau_{3,3})+a_{4,4} \left(-2 a_{5,5}\tau_{3,1}+2a_{5,4}\tau_{3,2}-a_{4,3}\tau_{3,3}+\tau_{3,3}^2\right)}{2 \left(a_{4,4}^3+3a_{4,4}^2 a_{5,4}+2 a_{5,5} (-a_{4,3}a_{5,4}+a_{4,2} a_{5,5}-a_{5,5}\tau_{3,2}+a_{5,4}\tau_{3,3})+a_{4,4} \left(2 a_{5,4}^2+a_{5,5} (2\tau_{3,3}-3 a_{4,3}-2 a_{5,3})\right)\right)}\,r^2\notag\\
&+\dfrac{\mu}{6 \left(a_{4,4}^3+3 a_{4,4}^2 a_{5,4}+2 a_{5,5} (-a_{4,3}a_{5,4}+a_{4,2}a_{5,5}-a_{5,5}\tau_{3,2}+a_{5,4} \tau_{3,3})+a_{4,4} \left(2a_{5,4}^2+a_{5,5} (-3a_{4,3}-2a_{5,3}+2\tau_{3,3})\right)\right)}\,r^3,
\end{align*}
where
\begin{align}
\mu&=2a_{4,4}^2\tau_{3,1}+2a_{4,4}a_{5,4}\tau_{3,1}-2 a_{4,4} (a_{4,3}+a_{5,3})\tau_{3,2}
-2 a_{5,5} (a_{4,3} \tau_{3,1}+\tau_{3,2} (-a_{4,2}+\tau_{3,2}))\notag\\
&+(2a_{4,3} (a_{4,3}+a_{5,3})-2a_{4,2} (a_{4,4}+a_{5,4})+2a_{5,5}\tau_{3,1}+3 a_{4,4} \tau_{3,2}+2 a_{5,4} \tau_{3,2}) \tau_{3,3}-(3a_{4,3}+2 a_{5,3}) \tau_{3,3}^2+\tau_{3,3}^3.\end{align}
 where $c_n=0$ for $n\geq 4$, 
we must have
$
\tau_{3,0}=6a_{5,0}+3a_{4,0}$ 
along with the vanishing of the three $4\times 4$-determinants, simultaneously,
\begin{align}\label{Case3_conf}
&\left|\begin{array}{cccc}
-\tau_{3,3}&a_{4,4}&2a_{5,5}&0\\
-\tau_{3,2}&a_{4,3}-\tau_{3,3}&2a_{5,4}+2a_{4,4}&6a_{5,5}\\
-\tau_{3,1}&a_{4,2} - \tau_{3,2}& 2a_{4,3} + 2a_{5,3} - \tau_{3,3}&3a_{4,4} + 6a_{5,4}\\
-6a_{5,0} - 3a_{4,0}& a_{4,1} - \tau_{3,1}& 2a_{4,2} + 2a_{5,2} - \tau_{3,2}& 
3a_{4,3} + 6 a_{5,3} - \tau_{3,3}\\
\end{array}\right|=0,\notag\\ \notag\\
&\left|\begin{array}{cccc}
-\tau_{3,3}&a_{4,4}&2a_{5,5}&0\\
-\tau_{3,2}&a_{4,3}-\tau_{3,3}&2a_{5,4}+2a_{4,4}&6a_{5,5}\\
-\tau_{3,1}&a_{4,2} - \tau_{3,2}& 2a_{4,3} + 2a_{5,3} - \tau_{3,3}&3a_{4,4} + 6a_{5,4}\\
0&- 6a_{5,0} -2 a_{4,0}& 2a_{4,1} + 2a_{5,1} - \tau_{3,1} & 3a_{4,2} + 6a_{5,2} - \tau_{3,2}\\
\end{array}\right|=0,\notag\\
\end{align}
and
\begin{align}
\left|\begin{array}{cccc}
-\tau_{3,3}&a_{4,4}&2a_{5,5}&0\\
-\tau_{3,2}&a_{4,3}-\tau_{3,3}&2a_{5,4}+2a_{4,4}&6a_{5,5}\\
-\tau_{3,1}&a_{4,2} - \tau_{3,2}& 2a_{4,3} + 2a_{5,3} - \tau_{3,3}&3a_{4,4} + 6a_{5,4}\\
0&0&- a_{4,0}-4a_{5,0}& 3a_{4,1} + 6a_{5,1} - \tau_{3,1}\\
\end{array}\right|=0,\notag\\
\end{align} 
and so on, for higher-order polynomial solutions. The vanishing of these determinants can be regarded as the conditions under which the coefficients $\tau_{3,1}$, $\tau_{3,2}$ and $\tau_{3,3}$ of Eq.(\ref{DE543}) are determined in terms of the other coefficients.
\end{theorem}
\vskip0.1true in
\begin{proof}  The proof of this theorem is rather lengthy: it employs the asymptotic iteration method in a similar way to the approach used by Saad {\it et al} in (2014) (\cite{saad2013}, Appendix A).
\end{proof}
\noindent We shall first verify the conclusions of this theorem regarding equation \eqref{DE_conf} by using the asymptotic iteration method followed by an analysis of the solutions for arbitrary parameters. To this end, we employ AIM for  \eqref{DE_conf} using
\begin{equation}\label{aim_conf}
\lambda_0=\frac{c}{b}+\frac{1-k}{r}+2 b r-\frac{2}{r-R},\quad s_0=b (2+k)-\frac{c^2}{4 b^2}-E+\dfrac{a}{\beta+r}+\frac{(k-1) (2 b+c\,R)}{2 bR\, r}+\frac{b(1- k)+R(c +2 b^2 R)}{b \, R\, (r-R)},
\end{equation}
and by means of 
\begin{align}\label{coeff}
a_{5,0}&=a_{5,1}=a_{5,5}=0,\quad a_{5,2}=-4b^2,\quad a_{5,3}=4b^2(R-\beta),\quad a_{5,2}=4b^2\beta\,R\notag\\
a_{4,0}&=8b^3,\quad a_{4,1}=4 b (c + 2 b^2 (\beta - R)),\quad a_{4,2}=-4 b (b - \beta c + b k + 2 b^2 \beta R + c R)\notag\\
a_{4,3}&=-4 b (\beta c R + b (\beta + \beta k + R - k R)),\quad a_{4,4}= 4\, b^2 \beta\, (k-1)\, R\notag\\
\tau_{3,0}&= -(4 b^3 (2 + k)-c^2 - 4 b^2 E),\quad \tau_{3,1}=-(4 a b^2 + 2 b c (1 + k) + (c^2 + 4 b^2 E) (R-\beta) + 
 4 b^3 (\beta (k+2) - k R)),\notag\\
\tau_{3,2}&=-(2 b (\beta\, c\, (1 + k)-2 b\, (k-1) )+(\beta\, (c^2 + 4 b^2 E)-4 a b^2  - 2 b c (k-1) - 4 b^3 \beta k)R),\quad \tau_{3,3}=2 b \beta (k-1) (2 b + c R),
\end{align}
 the necessary condition for the existence of polynomial solutions $f_n(r)=\sum_{k=0}^n c_k r^k$ of Eq. \eqref{DE_conf} becomes
\begin{equation}\label{En_aim_conf}
E_{n\ell}^d=b\, (2n'+k+2)-\dfrac{c^2}{4b^2},\qquad k=d+2l,
\end{equation}
where $n'$ refers to the degree of the polynomial solution of equation \eqref{DE_conf}  and is not necessarily equal to the number of nodes $n$ of the wave function. It is clear from \eqref{coeff}, there is no zero-degree polynomial solution available.  For the first-degree polynomial solution, we have
\begin{equation}\label{En_aim_conf_case1}
E_{n\ell}^d=b\, (4+k)-\dfrac{c^2}{4b^2},\quad f_1(r)=1+\left(\frac{c}{2 b}+\frac{1}{R}\right)\,r
\end{equation}
providing
\begin{align}\label{exact_conf_cond}
4 a b^2+2 b c (3+k)+\left(2 a b c+c^2 (3+k)\right) R+4 b^2\, c\, R^2&=0\notag\\
2 b (\beta c (3+k)-4 b k)+c (\beta c (3+k)-4 b k)R+\left(16 b^3-2 a b c+4 b^2\beta c-c^2 (1+k)\right)R^2&=0\notag\\
8 b^2\, \beta\, k+4 b\, \beta\, c\, k\, R+\left(4 a\, b^2+\beta \left(c^2 (1+k)-16\, b^3\right)\right)R^2&=0.
\end{align}
In Table \ref{table:difdn6}, we report the exact eigenvalues $E_{00}^{3}=7b-c^2/(4b^2)$ using the roots of the equations given by \eqref{exact_conf_cond} and the results obtained by AIM initiated with $r_0=R/2$ for different values of $R$ and $\beta$, where we have fixed $k=3$. For arbitrary values of the potential parameters, we can employ AIM initiated with \eqref{aim_conf} to obtain accurate eigenvalues as the roots of the termination condition \ref{tm_cond}. Some of the these results are reported in Table \ref{table:difdn6}. There is an interesting additional application of AIM for these confining potentials: it is possible to use the termination condition to find the proper radius of confinement $R$ for a particular energy; in other words, we may regard the termination condition as function of $(r,R)$ given a particular energy $E$. Consider for example $\beta=a=b=-c=1, k=3$, and $E=9$, what is the radius of confinement for this particular case? The direct application of AIM implies that $R=1.074~414~209~270~221~205$ while for $E=10$, the proper radius of confinement $R=1.016~954~256~339~063~400.$ The method can be easily generalized for arbitrary values of the parameters.

\begin{table}[!h] \caption{A comparison between selected eigenenergies calculated using AIM with the exact values obtained as the roots of equation \eqref{exact_conf_cond}.\\ } 
\centering 
\begin{tabular}{|c|c|c|c|c|p{1.7in}|c|| }
\hline
$a$&$b$&$c$&$\beta$&$R$&$E_{AIM}$& $E_{exact}$\\ \hline
${20}/{3}$&${11}/{6}$&$-{11}/{6}$&2&1&$12.583~333~333~333~333~333_{N=3}$&$151/12$\\ \hline
${28}/{15}$&${13}/{30}$&$-{13}/{90}$&3&2&$~3.005~555~555~555~555~556_{N=3}$&$541/180$\\ \hline
${55}/{63}$&${47}/{252}$&$-{47}/{1512}$&4&3&$~1.298~611~111~111~111~111_{N=3}$&$187/144$\\ \hline
${143}/{90}$&${71}/{360}$&$-{71}/{1350}$&5&3&$~1.362~777~777~777~777~778_{N=3}$&$2453/1800$\\ \hline
${91}/{180}$&${37}/{360}$&$-{37}/{3600}$&5&4&$~0.716~944~444~444~444~444_{N=3}$&$2581/3600$\\ \hline
${14}/{15}$&${13}/{120}$&$-{13}/{720}$&6&4&$~0.751~388~888~888~888~889_{N=3}$&$541/720$\\ \hline
 \hline
\end{tabular}
\label{table:difdn6}
\end{table}

\begin{table}[h] \caption{Eigenvalues $E_{n\ell}^{d=3}$ for $V(r)=1/(r+1)-r+r^2$ for different radius of confinement $R$ . The initial value employed by AIM is $r_0=R/2$. The subscript $N$ refers to the number of iterations used by AIM.\\ } 
\centering 
\begin{tabular}{|c|c |p{2.5in}||c|c| p{2.5in}|}
\hline
$R$&$n$&$E_{n0}^{d=3}$&$R$&$n$&$E_{n0}^{d=3}$\\ \hline
$1$&0&$~10.328~716~871~106~505~751_{N=36}$&$2$&0&$~3.105~413~452~488~593~322_{N=56}$\\
~&1&$~39.987~716~212~123~541~087_{N=37}$&~&1&$10.692~851~715~920~035~023_{N=55}$\\
~&2&$~89.345~269~629~504~444~833_{N=36}$&~&2&$23.063~954~484~826~017~705_{N=57}$\\
~&3&$158.435~845~294~778~224~568_{N=41}$&~&3& $40.347~069~688~147~624~366_{N=55}$ \\
 \hline
\hline
$R$&$\ell$&$E_{0\ell}^{d=3}$&$R$&$\ell$&$E_{0\ell}^{d=3}$\\ \hline
$1$&0&$10.328~716~871~106~505~751_{N=36}$&$2$&0&$~3.105~413~452~488~593~322_{N=56}$\\
~&1&$20.608~236~713~301~997~322_{N=36}$&~&1&$~5.819~309~536~633~945~722_{N=55}$\\
~&2&$33.620~107~194~959~911~851_{N=36}$&~&2&$~9.196~161~676~541~214~605_{N=56}$\\
~&3&$49.228~314~838~693~690~037_{N=36}$&~&3& $21.521~551~806~858~223~355_{N=59}$ \\
 \hline
\hline
\end{tabular}
\label{table:difdn6}
\end{table}
\section{Conclusion}\label{conc}

\noindent In this work exact and approximate solutions of Schr\"odinger's equation with softcore Coulomb potentials under hard and soft confinement were found. These problems generate an interesting class of differential equation that goes beyond the classical problems which have solutions of hypergeometric type.  In this paper the problems were  analyzed as special cases of a very general scheme for the study of linear second-order differential equations with polynomial coefficients that admit polynomial solutions.  Necessary and sufficient conditions are derived for the existence of such solutions. The methods presented in this work allow us to obtain compact algebraic  expressions for the exact analytical solutions.  These are then verified by the asymptotic iteration method. In cases where the parametric conditions for exact polynomial solutions are  not met, the asymptotic iteration method is employed directly to find highly accurate numerical solutions. In this work, the asymptotic iteration method served two purposes.  The first was to confirm the validity of the sufficient conditions obtained analytically. The second is to provide approximate solutions to the eigenvalue problems, whether potential parameters are specially restricted or freely chosen. For both purposes, the method proves to be extremely effective and provides very accurate results. It is also clear from the present work that the method and the analytic expressions obtained for the different classes of the differential equations can be easily adapted to study other eigenproblems appearing in theoretical physics.

\section{Acknowledgments}
\medskip
\noindent Partial financial support of this work under Grant Nos. GP3438 and GP249507 from the 
Natural Sciences and Engineering Research Council of Canada
 is gratefully acknowledged by us (respectively RLH and NS). 

\end{document}